\documentclass[aps,amsfonts,superscriptaddress]{revtex4}

\newcommand{\be}{\begin{equation}}
\newcommand{\ee}{\end{equation}}
\newcommand{\beqn}{\begin{eqnarray}}
\newcommand{\eeqn}{\end{eqnarray}}
\newcommand{\Eq}[1]{Eq.~(\ref{#1})}
\newcommand{\Eqs}[1]{Eqs.~(\ref{#1})}

\date{August 2002}
\begin{document}

\title{A General Approach of Quasi-Exactly Solvable Schr\" odinger 
Equations with Three Known Eigenstates}
\author{N. DEBERGH}
\email[Chercheur, Institut Interuniversitaire des
Sciences Nucl\'eaires, Bruxelles ]{E-mail: Nathalie.Debergh@ulg.ac.be} 
\affiliation{\it Fundamental Theoretical Physics,
Institute of Physics (B5),
University of Li\`ege,
B-4000 LIEGE  (Belgium)}
\author{J. NDIMUBANDI}
\email{jndimubandi@yahoo.fr}
\affiliation{University of Burundi, Department of Mathematics, 
P.O. Box 2700, BUJUMBURA (Burundi)}
\author{B. VAN DEN BOSSCHE}
\email{bvandenbossche@ulg.ac.be}
\affiliation{\it Fundamental Theoretical Physics,
Institute of Physics (B5),
University of Li\`ege,
B-4000 LIEGE  (Belgium)}

\begin{abstract}
We propose a general method for constructing quasi-exactly solvable
 potentials with three analytic eigenstates. These potentials can 
be real or complex functions but the spectrum is real. A comparison
 with other methods is also performed.
\end{abstract}

\maketitle

\newpage

\section{Introduction}
\label{Section1}

Quasi-exactly solvable (Q.E.S.) Schr\" odinger equations are of great 
current interest. Indeed they have the characteristic property that 
analytic expressions for a finite number of energy levels and their 
corresponding eigenfunctions can be obtained. These Q.E.S. equations 
have been first investigated in \cite{Ref1} and since then they have 
been studied extensively (for a detailed review see f.i. \cite{Ref2}). 
Actually one can find two different points of view for analyzing these 
equations.

First, the number of analytic solutions is related \cite{Ref3} 
to the dimension of the irreducible representations of the Lie 
algebra $sl(2,R)$. More specifically the corresponding Schr\" odinger 
Hamiltonians can be written, up to a change of variables and a 
change of eigenfunctions, as linear and quadratic combinations 
of the $sl(2,R)$ generators, these preserving the finite-dimensional 
space of the solutions.

Second, the number of analytic solutions is fixed at the 
start (essentially two \cite{Ref4} or three \cite{Ref5}). 
Except in the case of two solutions \cite{Ref6}, the corresponding 
Schr\" odinger Hamiltonians cannot, in general, be related to any 
Lie algebra.

Each of these two points of view has its own advantages and disadvantages. 
The Lie algebraic approach is elegant, straightforward and the 
associated potentials can be treated in a global way. 
However it concerns only a few interactions which have been 
listed in \cite{Ref3}. The second approach deals with a bigger 
 and not yet known number of potentials but no general method to study 
them is available. Until now, the only method found in the 
literature to construct these Q.E.S. potentials is 
the one developed in \cite{Ref4} and \cite{Ref5}. It is based on a 
recursive construction of a finite number of supersymmetric
 partners \cite{Ref7}. It will be recalled in Section~\ref{Section4} 
with more details but let us already mention 
that the Kuliy-Tkachuk method is too constraining and thus 
misses some Q.E.S. interactions.  Some of them are given in the following.

The purpose of this paper is then to propose an alternative to 
this method in order to approach the Q.E.S. equations with a 
known number (we will concentrate essentially on three) 
of analytic solutions in a general way. To do so we will make use 
of the general algorithm developed in \cite{Ref8}.

The paper is organized as follows: In Section~\ref{Section2}, we explain 
the formalism of our method. It is illustrated through examples in 
Section~\ref{Section3}. In Section~\ref{Section4}, we compare our 
method with the one of Kuliy and Tkachuk. Finally, we draw
our conclusions in Section~\ref{Section5}.

\section{The general approach}
\label{Section2}

We concentrate on the one-dimensional and time-independent 
Schr\" odinger equation
\be
\left[-\frac{d^2}{dx^2}+V(x)\right]\psi_N(x)=E_N \psi_N(x).
\label{Eq1}
\ee
The eigenfunctions $\psi_N(x)$ as explained in \cite{Ref8} are written as
\be
\psi_N(x)=g(x) \left[f(x)\right]^{\lambda_N} 
\sum_{m=0}^N c_m^{(N)} \left[h(x)\right]^m
\label{Eq2}
\ee
where the function $g(x)$ is a weight factor, $[f(x)]^{\lambda_N}$ 
is introduced due to eventual singularities in $V(x)$, 
and the quantities (real or complex) 
$c_m^{(N)}$ are expansion coefficients on the basis $h(x)$. 
Actually this basis plays the prominent role in our approach in the sense 
that each of the functions entering our formalism can be expanded 
in this basis according to
\beqn
g'(x)&=&-g(x)\sum_{l=0}^M g_l^1 \left[h(x)\right]^l,
\label{Eq3}\\
f(x)&=&\sum_{l=0}^M f_l^0 \left[h(x)\right]^l,
\label{Eq4}\\
f'(x)&=&\sum_{l=0}^M f_l^1 \left[h(x)\right]^l,
\label{Eq5}\\
h'(x)&=&\sum_{l=0}^M h_l^1 \left[h(x)\right]^l,
\label{Eq6}
\eeqn
the prime standing for the derivative with respect to $x$. 
The upper-right index for the coefficients 
$g_l^1, f_l^0, f_l^1,  h_l^1$ indicates the derivative order of 
the corresponding expanded function. The upper summation index $M$ is 
in general an arbitrary positive integer, big enough to be the highest 
value of all upper summation indexes in \Eqs{Eq3}--(\ref{Eq6}), 
the added expansion coefficients being vanishing when necessary. 
Finally the upper summation index $N$ is also a positive integer 
left arbitrary until we consider specific examples in the next Section.

In \cite{Ref8}, we have limited ourselves to specific potentials of the form
\be
V(x)=\sum_{k=-V_{min}}^{V_{max}} V_k \left[f(x)\right]^k 
\sum_l v_{l,k} \left[h(x)\right]^l
\label{Eq7}
\ee
the coefficients $V_k$ and $v_{l,k}$ being fixed according to the 
interaction we want to deal with while $l$ runs in the space of integers. 
The positive integers $V_{min}$ and $V_{max}$ are also determined 
once the interaction is given. 
The expansion~(\ref{Eq7}) has been proposed to include all known 
Q.E.S. potentials. However, in the present paper, we do not have to 
ask that the potential obeys a given expansion 
due to the fact that the number of eigenfunctions is fixed at the start.
(We will nevertheless see in the next Section
that the determined potential will be of the
type of \Eq{Eq8}.)

More precisely, if we consider three eigenstates as stated 
in the Introduction, we have two possibilities: 
Either these eigenfunctions are of the form given in \Eq{Eq2} 
with $N=L, L+1, L+2$, or they coincide with  \Eq{Eq2} with the same fixed 
$N=L$, but different values of $c_m^{(N)}$ in order to have linearly 
independent polynomials in the basis $h(x)$. 
The analysis of already known Q.E.S. potentials reveals that 
the Kuliy-Tkachuk approach is included in the first possibility 
while the Lie algebraic one can be found in the second possibility.

In both cases, we have the first eigenstate equal to
\be
\psi_L(x)= g(x) \left[f(x)\right]^{\lambda_L} 
\sum_{m=0}^L c_m^{(L)} \left[h(x)\right]^m
\label{Eq8}
\ee
and, for simplicity, associated to the energy 
\be
E_L=0.
\label{Eq9}
\ee
Such a requirement determines the potential $V(x)$ up to 
the presence of parameters which will be fixed asking for the 
existence of the other solutions. 
More precisely, by inserting \Eqs{Eq8} and~(\ref{Eq9}) in \Eq{Eq1} 
and using \Eqs{Eq3}--(\ref{Eq6}), the potential is
\beqn
&&V(x)=\left\{f_p^0 f_q^0 c_m^{(L)} \left[h(x)\right]^{p+q+m}\right\}^{-1} 
\bigg\{f_k^0 f_l^0 g_m^1 g_n^1 c_s^{(L)}\left[h(x)\right]^{k+l+m+n+s} 
\nonumber \\
&&\mbox{}-m f_k^0 f_l^0 g_m^1 h_n^1 
c_s^{(L)} \left[h(x)\right]^{k+l+m+n+s-1} 
+ \lambda_L l f_l^1 h_k^1 f_m^0 c_s^{(L)} \left[h(x)\right]^{k+l+m+s-1}
\nonumber \\
&&
\mbox{}+(\lambda_L^2-\lambda_L) f_k^1 f_l^1 c_s^{(L)}
\left[h(x)\right]^{k+l+s} 
-2\lambda_L g_k^1 f_l^0 f_m^1 c_s^{(L)}\left[h(x)\right]^{k+l+m+s}\nonumber \\
&&\mbox{}+2\lambda_L s f_k^0 f_l^1 h_m^1 c_s^{(L)} 
\left[h(x)\right]^{k+l+m+s-1}
-2s f_k^0 f_l^0 g_m^1 h_n^1 c_s^{(L)} \left[h(x)\right]^{k+l+m+n+s-1} 
\nonumber \\
&&\mbox{}+f_k^0 f_l^0 h_m^1 h_n^1 c_s^{(L)}
s(s-1+m)\left[h(x)\right]^{k+l+m+n+s-2}\bigg\}.
\label{Eq10}
\eeqn
In this expression and the following ones, we have, for simplicity, omitted 
all the summation indexes but it is understood that there is a summation 
on each repeated index (except $L$).
 
Now the potential of \Eq{Eq10} has to be constrained so that there exist 
other solutions to it. For the first possibility mentioned above, i.e.,
the one including in particular the Kuliy-Tkachuk potentials, we just 
have to require that the functions~(\ref{Eq2}) with $N=L+1$ and $N=L+2$ 
are solutions of \Eq{Eq1} with $V(x)$ fixed according to \Eq{Eq10}. 
This gives rise to (two) systems of algebraic equations 
(instead of differential ones). This is the main advantage, 
besides its generality, of our approach based on the development of all 
functions in the $h(x)$-basis. These systems are specified by 
vanishing the different coefficients of linearly independent terms in
\begin{eqnarray}
&&-f_k^0 f_l^0 h_m^1 h_n^1 c_s^{(L)}c_t^{(N)}t(t-1+m)
\left[h(x)\right]^{k+l+m+n+s+t-2}
+2t f_k^0 f_l^0 g_m^1 h_n^1 c_s^{(L)}c_t^{(N)}\left[h(x)\right]^{k+l+m+n+s+t-1}
\nonumber \\
&&\mbox{}-2 \lambda_N f_k^0 f_l^1 t h_m^1 c_s^{(L)} c_t^{(N)}
\left[h(x)\right]^{k+l+m+s+t-1}+(\lambda_L -\lambda_N) 
\bigg\{f_k^0 f_l^1 h_m^1 l c_s^{(L)} c_t^{(N)}\left[h(x)\right]^{k+l+m+s+t-1} 
\nonumber \\
&&\mbox{}-2f_k^0 g_l^1 f_m^1 c_s^{(L)}c_t^{(N)}\left[h(x)\right]^{k+l+m+s+t}
+(\lambda_L + \lambda_N -1)f_k^1 f_l^1 c_s^{(L)}c_t^{(N)}
\left[h(x)\right]^{k+l+s+t}\bigg\}\nonumber \\
&&\mbox{}+2 \lambda_L f_k^0 f_l^1 h_m^1 s c_s^{(L)}c_t^{(N)}
\left[h(x)\right]^{k+l+m+s+t-1}
-2f_k^0 f_l^0 g_m^1 s h_n^1 c_s^{(L)}c_t^{(N)}
\left[h(x)\right]^{k+l+m+n+s+t-1}\nonumber \\
&&\mbox{}+f_k^0 f_l^0 h_m^1 h_n^1 s(s-1+m)c_s^{(L)} 
c_t^{(N)}\left[h(x)\right]^{k+l+m+n+s+t-2} 
-E_N f_k^0 f_l^0 c_s^{(L)}c_t^{(N)}(h(x))^{k+l+s+t}=0,
\label{Eq11}
\end{eqnarray}
where $N=L+1, L+2$.
It is clear that if we want to consider the case of two only eigenstates, 
we have to solve only one system, the one corresponding to $N=L+1$. 
We can also extend this approach to the case of more than three solutions. 
The potential  of \Eq{Eq10} is then more and more constrained 
until it is impossible to fulfill the constraints. 

In what concerns the second possibility, i.e., the one associated 
to the Lie algebraic approach, we have to ask for the function in \Eq{Eq2} 
to be a solution of \Eq{Eq1} with $N=L$, together with other coefficients 
than $c_m^{(L)}$, say $\tilde{c}_m^{(L)}$. 
The system of constraints is now:
\beqn
&&-f_k^0 h_m^1 h_n^1 c_s^{(L)}\tilde{c}_t^{(L)}(t-s)(t-1+m+s)
\left[h(x)\right]^{k+m+n+s+t-2} 
+2f_k^0 g_m^1 h_n^1 c_s^{(L)}\tilde{c}_t^{(L)}(t-s)
\left[h(x)\right]^{k+m+n+s+t-1} \nonumber \\
&&\mbox{}-2 \lambda_L f_k^1 h_n^1 c_s^{(L)}
\tilde{c}_t^{(L)}(t-s) \left[h(x)\right]^{k+n+s+t-1}=0.
\label{Eq12}
\eeqn
The number of independent sets of coefficients $\tilde{c}_t^{(L)}$ 
satisfying \Eq{Eq12} gives the number of eigenstates, besides $\psi_L(x)$
given in \Eq{Eq8}.

\section{Examples}
\label{Section3}

Let us now illustrate our approach through specific examples. 
The point is simply to choose the function $f(x)$ in an arbitrary way 
and then to adjust the important basis function 
$h(x)$ so that \Eqs{Eq4}--(\ref{Eq6}) are satisfied. 
Our recent analysis of Q.E.S. equations in \cite{Ref8} has revealed that 
the choice
\be
f(x)=1+x^2
\label{Eq13}
\ee
covers many Q.E.S. interactions. 
This, as well as $f'(x)=2x$, clearly suggests the basis function $h(x)=x$. 
From \Eqs{Eq4}--(\ref{Eq6}), we get the non vanishing coefficients
\be
f_0^0=1, f_2^0=1, f_1^1=2, h_0^1=1.
\label{Eq14}
\ee
For the first possibility, corresponding to the constraints in \Eq{Eq11}, 
let us fix $L=1$ and choose, without loss of generality, $c_1^{(1)}=1$. 
The potential of \Eq{Eq10} then reads
\beqn
V(x)&=&g_m^1 g_n^1 x^{m+n}-mg_m^1 x^{m-1}+
\frac{4 \lambda_1^2-2 \lambda_1}{1+x^2}
-\frac{4 \lambda_1^2-4 \lambda_1}{(1+x^2)^2}
-\frac{4 \lambda_1}{1+x^2} g_m^1 x^{m+1} \nonumber \\
&&
\mbox{}+\frac{4 \lambda_1 x}{(1+x^2)(c_0^{(1)}+x)}
-2\frac{g_m^1 x^m}{c_0^{(1)}+x}.
\label{Eq15}
\eeqn
A rapid look at the first constraint,  corresponding to $N=2$ in \Eq{Eq11}, 
shows that the running index $m$ has in fact to be restricted to the 
values $m=0$ and $m=1$. This first constraint  
is then a system of eight (nonlinear) equations 
for the eight unknowns $g_0^1, g_1^1,  c_0^{(1)},c_0^{(2)}, c_1^{(2)},
\lambda_1,\lambda_2, E_2$ if, once again, we take advantage of the fact 
that the eigenstates are fixed up to a normalization constant, 
so that $c_2^{(2)}$ can be chosen equal to 1. 
The second constraint, corresponding to $N=3$, is a system of nine 
(nonlinear) equations for the nine unknowns 
$g_0^1, g_1^1, c_0^{(1)}, c_0^{(3)}, c_1^{(3)}, c_2^{(3)}, 
\lambda_1, \lambda_3, E_3$ with, for the same reason, $c_3^{(3)}=1$. 
Thus there is a total of seventeen equations for thirteen unknowns. 
Nevertheless, this system is rather easy to handle and furnishes 
two sets of solutions. The first set is the one corresponding to 
the well-known harmonic oscillator potential while the second one is
\beqn
&&g_0^1=0, g_1^1=\sqrt{c}, \lambda_1=\lambda_2=\sqrt{c}-\frac{1}{2}, 
\lambda_3=\frac{3}{2}-\sqrt{c}, \nonumber \\ \nonumber \\
&&E_2=2\sqrt{c}, E_3=-8c+12\sqrt{c}, c_0^{(1)}=0, c_0^{(2)}=-1, c_1^{(2)}=0, 
\nonumber \\ \nonumber \\
&&c_0^{(3)}=0, c_1^{(3)}=\frac{1}{\frac{4}{3}c-2}, c_2^{(3)}=0
\label{Eq16}
\eeqn
with $c$ being
\be
c=\left\{-1+\sqrt{5}\cos\left[\frac{1}{3} 
\arctan\left(\frac{\sqrt{109}}{4}\right)\right]\right\}
^2\label{Eq17},
%&=&\frac{7}{2}-\frac{\sqrt{73}}{4}\cos\left[\frac{1}{3} 
%\arctan\left(\frac{12\sqrt{109}}{611}\right)\right]
%-\frac{\sqrt{219}}{4}\sin\left[\frac{1}{3} 
%\arctan\left(\frac{12\sqrt{109}}{611}\right)\right],
%\label{Eq17snd}
\ee
i.e., $c=1.119639$. The associated potential is, from  \Eq{Eq10}, 
\be
V(x)=c x^2-4c-\sqrt{c}+\frac{8c-4\sqrt{c}}{1+x^2}-
\frac{4c-8\sqrt{c}+3}{(1+x^2)^2}
\label{Eq18}
\ee
and is, to our knowledge, a new Q.E.S. potential. 
It admits three analytic (and physical) solutions given by
\beqn
\psi_1(x)&=&x(1+x^2)^{\sqrt{c}-\frac{1}{2}} 
\exp\left(-\frac{\sqrt{c}x^2}{2}\right), \; E_1=0,
\label{Eq19}\\
\psi_2(x)&=&(1-x^2)(1+x^2)^{\sqrt{c}-\frac{1}{2}} 
\exp\left(-\frac{\sqrt{c}x^2}{2}\right), \; E_2=2\sqrt{c},
\label{Eq20}\\
\psi_3(x)&=&x\left[1-\left(2-\frac{4}{3}c\right)x^2\right]
(1+x^2)^{-\sqrt{c}+\frac{3}{2}} 
\exp\left(-\frac{\sqrt{c}x^2}{2}\right), \; E_3=-8c+12\sqrt{c}.
\label{Eq21}
\eeqn
We remark that the same analysis, but with $L=0$, 
would have led to the potential given in \cite{Ref5}. Thus, our method
includes the one described in that reference. Moreover, as we will show in
the next Section, it is  simpler (see the function $U(x)$ of \Eq{Eq39}
which would be needed to find the potential \Eq{Eq18} and the eigenfunctions
\Eqs{Eq19}--(\ref{Eq21})).

For the second possibility, corresponding to the 
constraints in \Eq{Eq12}, let us fix $L=2$, which is the minimal value 
for having at most three eigenstates in this case, together 
with $c_L^{(L)}=1$ and $\tilde{c}_L^{(L)}=1$. 
We have not found a new potential with three eigenstates. Only known ones 
like the harmonic oscillator or its supersymmetric partner \cite{Ref5} emerge.
We have however found a new potential with two eigenstates. It is given by
\be
V(x)=\frac{x^2}{36}-\frac{1}{6}+\frac{4+2ix}{3(x+i)^2}
\label{Eq22}
\ee
and the two eigenstates and related eigenvalues are
\beqn
\psi_2(x)&=&(3+2ix+x^2)(1+x^2)^{-1}
\exp\left(-\frac{x^2}{12}\right), \; E_2=0,
\label{Eq23}\\
\tilde{\psi}_2(x)&=&(-1+2ix+x^2)
\exp\left(-\frac{x^2}{12}\right), \; \tilde E_2=\frac{2}{3}.
\label{Eq24}
\eeqn
Thus, we have here an example of a complex but PT-symmetric \cite{Ref9} 
Q.E.S. potential, with real eigenvalues. 
This potential is new compared to the ones already studied through other 
developments \cite{Ref10}.

To end this Section, we mention that the potentials \Eq{Eq18} and \Eq{Eq22}
could be written as in \Eq{Eq7}, implying that, given the
potentials, we could have found the eigenfunctions using the algorithm
we have described in \cite{Ref8}.

\section{Comparison with the Kuliy-Tkachuk approach}
\label{Section4}

For completeness, let us briefly recall what the method of  Kuliy-Tkachuk 
is. It is  based on a recursive construction of supersymmetric partners. 
Indeed, the  initial potential $V(x)$ is required to be written as
\be
 V(x) =  W_L(x)^2 - W'_L(x).
\label{Eq25}
\ee
If such a $W_L(x)$ does exist, then the ground state and its energy 
are known
\be
\psi_L(x)= \exp\left[-\int W_L(x)dx\right],\; E_L=0.
\label{Eq26}
\ee
Provided one can find two other functions $W_{L+1}(x)$ 
and $W_{L+2}(x)$ such that the Ricatti equations
\be
W_k^2(x) + W'_k(x) + E_k = W_{k+1}^2(x) - W'_{k+1}(x) + E_{k+1}, \; k=L,L+1
\label{Eq27}
\ee
are satisfied, two other solutions (respectively associated with $E_{L+1}$ 
and $E_{L+2}$) are available. They are given by
\beqn
\psi_{L+1}(x)&=&\left[-\frac{d}{dx}+W_L(x)\right]
\exp\left[-\int W_{L+1}(x)dx\right],
\label{Eq28}\\
\psi_{L+2}(x)&=&\left[-\frac{d}{dx}+W_L(x)\right]
\left[-\frac{d}{dx}+W_{L+1}(x)\right]
\exp\left[-\int W_{L+2}(x)dx\right].
\label{Eq29}
\eeqn
The main point of the Kuliy-Tkachuk approach is to reduce 
\Eqs{Eq27} to only one equation
\beqn
&&W_+^{(L)}(x)W_+^{(L+1)}(x)\left[W_+^{(L+1)}(x)-W_+^{(L)}(x)\right]
-\frac{d}{dx}\left[W_+^{(L)}(x)W_+^{(L+1)}(x)\right]
\nonumber\\
&&\hspace{3.5cm}\mbox{}+\left(E_{L+2}-E_{L+1}\right)
W_+^{(L)}(x)+E_{L+1}W_+^{(L+1)}(x)=0
\label{Eq30}
\eeqn
of the two functions
\be
W_+^{(k)}(x) \equiv W_k(x) + W_{k+1}(x), \; k=L,L+1.
\label{Eq31}
\ee
Finally, \Eq{Eq30} can be considered as an algebraic one in terms 
of $W_+^{(L)}(x)$, for instance, if the function
\be
U(x) \equiv W_+^{(L)}(x) W_+^{(L+1)}(x)
\label{Eq32}
\ee
is fixed arbitrarily at the start. 
The whole approach thus relies on an 
adequate choice of this function $U(x)$. Note that the authors of 
Ref.~\cite{Ref5} have restrained themselves to the context of the 
functions $U(x)$ such that
\beqn
U(x) &\leq 0& {\rm if}\; x \in [{\rm min} (x_0,\tilde x_0),{\rm max} 
(x_0,\tilde x_0)],
\label{Eq33}\\
U(x) &> 0& {\rm otherwise}
\label{Eq34}
\eeqn
where $x_0$ and $\tilde x_0$ are the two zeroes of $U(x)$ such that
\be
U'({\rm min} (x_0,\tilde x_0)) < 0,\quad U'({\rm max} (x_0,\tilde x_0)) > 0.
\label{Eq35}
\ee
In~\cite{Ref5}, these conditions were fixed according to the 
requirement that none of 
the functions $W_l(x)$, $l=L, L+1, L+2$ possesses singularities. 

Now we concentrate on the determination of this $U(x)$ related to the 
examples developed in the previous Section.

In what concerns the potential \Eq{Eq18} and its 
eigenstates~(\ref{Eq19})--(\ref{Eq21}), a 
comparison with \Eqs{Eq26} and~(\ref{Eq28})--(\ref{Eq29}) 
gives rise to the identifications
\beqn
W_L(x)&=&\sqrt{c} x -\frac{(2\sqrt{c}-1)x}{(1+x^2)}-\frac{1}{x},
\label{Eq36}
\\
W_{L+1}(x)&=& \sqrt{c} x - \frac{(2\sqrt{c}+1)x}{(1+x^2)}+\frac{1}{x},
\label{Eq37}\\
W_{L+2}(x)&=&\sqrt{c} x +\frac{(2\sqrt{c}+1)x}{(1+x^2)}-\frac{1}{x} 
-\frac{2x
\left[(14c-3\sqrt{c}-15)+(68c+12\sqrt{c}-90)x^2\right]}{(34c+6\sqrt{c}-45)
x^4+(14c-3\sqrt{c}-15)x^2-3\sqrt{c}}.
\label{Eq38}
\eeqn
It is already evident that such functions will lead to a Q.E.S. potential 
excluded in the Kuliy-Tkachuk approach due to the singularity $x=0$ in 
\Eqs{Eq36}--(\ref{Eq38}). Indeed the corresponding $U(x)$ can be obtained 
through \Eqs{Eq31}--(\ref{Eq32}) and is
\be
U(x)=2\sqrt{c}x^2(x^2-1)
\frac{(-192c-39\sqrt{c}+255)x^2
+2(-17c+3\sqrt{c}+15)}{(34c+6\sqrt{c}-45)x^4
+(14c-3\sqrt{c}-15)x^2-3\sqrt{c}}.
\label{Eq39}
\ee
By opposition to the $U(x)$ permitted in \cite{Ref5}, this one admits 
one second-order zero point ($x=0$) and two ordinary zero points ($x=\pm 1$). 
Moreover $U(x)$ has no pole and is such that (compare with 
\Eqs{Eq33}--(\ref{Eq35}))
\beqn
U(x) &\leq 0&  {\rm if}\; x \in [-1,1],
\label{Eq40}\\
U(x) &> 0 &{\rm if}\;  x \in ]-\infty , -1[ \; \cup \; ]1,+\infty [,
\label{Eq41}\\
U'(0)&=0,&  U'(-1) < 0, \quad U'(1) > 0.
\label{Eq42}
\eeqn
Let us also stress the heaviness of the $U(x)$ given in \Eq{Eq39}: The least 
one can say is that it is not natural to think at the start of such a 
function while the choice of our function $f(x)=1+x^2$ is on the other 
hand quite straightforward. This is the main advantage of our approach:
using $L=0$ as a starting point in \Eq{Eq8}, we could have obtained
the Kuliy-Tkachuk solution of Ref.~\cite{Ref5} doing the same type of
easy calculations that we have performed here. We would have of course
obtained the same results as has been used in that reference, i.e.:
\beqn
V(x)&=&\frac{3}{4}x^2+\frac{1}{2}\left(6-7\sqrt{3}\right)
-\frac{2\left(-3+\sqrt{3}\right)}{1+x^2}
+\frac{2\left(-3+2\sqrt{3}\right)}{(1+x^2)^2},
\label{Eq421}\\
\psi_0(x)&=&(1+x^2)^{(3-\sqrt{3})/2}\exp\left(-\frac{\sqrt{3}}{4}x^2\right),
\; E_0=0,
\label{Eq422}\\
\psi_2(x)&=&x(1+x^2)^{(\sqrt{3}-1)/2}\exp\left(-\frac{\sqrt{3}}{4}x^2\right),
\; E_1=6-3\sqrt{3},\label{Eq423}\\
\psi_3(x)&=&(1-x^2)(1+x^2)^{(\sqrt{3}-1)/2}
\exp\left(-\frac{\sqrt{3}}{4}x^2\right),\; E_2=6-2\sqrt{3}.
\label{Eq424}
\eeqn
 On the
contrary, it is doubtful that the function $U(x)$ of \Eq{Eq39} could
have been witnessed from the start, while, in our approach, it just consists
in starting with $L=1$ in \Eq{Eq8}. It is also interesting to remark that
the potential $V(x)$ of~\cite{Ref5}, and our potential of \Eq{Eq18} have the
same type of $x$-dependence: $V(x)=a+bx^2+c/(1+x^2)+d/(1+x^2)^2$, and admit
three solutions. This shape of the potential is very rich, 
see e.g~\cite{Ref8}. In fact it is even possible 
(apart from the harmonic oscillator $c=d=0$) to find an exactly solvable
model, using $a=0, b=1/4,c=4,d=-8$. One may wonder why it is not just possible 
to study the Schr\"odinger equation with a potential of this general
shape. We have shown in Ref.~\cite{Ref8} that the constraints are
extremely difficult to handle, especially when the number of solutions is 
growing.

For the potential of \Eq{Eq22}, the function $U(x)$ simply reduces 
to $W_+^{(L)}(x)$ as defined in \Eq{Eq31} due to the fact that this 
potential has two eigenvalues only. We respectively have
\beqn
W_L(x)&=&\frac{x}{6}+\frac{1}{x+i}-\frac{1}{x+3i},
\label{Eq43}\\
W_{L+1}(x)&=&\frac{x}{6}-\frac{1}{x+i}+\frac{1}{x+3i}-\frac{1}{x+4i}
\label{Eq44}
\eeqn
evidently leading to
\be
U(x)\equiv W_+^{(L)}(x)=\frac{x}{3}-\frac{1}{x+4i}.
\label{Eq45}
\ee
So once again singularities (in the complex plane) arise and the 
resulting $U(x)$ could not be found following the method of Ref. \cite{Ref5}.

\vspace{0.5cm}

{\noindent \it Remarks concerning the choice of $U(x)$}

\vspace{0.2cm}

We have already seen that the choice of $U(x)$ is a non trivial task.
We want to show here that the choice is also sensitive to small
changes of parameters, and that care must be done to extract meaningful 
solutions, i.e., even if $U(x)$ allows to generate a well-behaved potential
$V(x)$, the eigenfunctions obtained using the formalism are not physically
suitable.
We show this starting with the simple choice $U(x)=cx^2$, with $c$ a 
constant to be determined. To avoid a ${\cal O}(1/x^2)$ 
singularity in the potential, one is forced to adjust the constant 
$c$ to $c=E_1(E_2-E_1)$,
with $E_0=0$. Choosing $E_1=1,E_2=2$, so as to have an equally spaced
spectrum, two potentials emerge: either $V(x)=(x^2+10)/4$, or 
$\tilde{V}(x)=(x^2-2)/4$, which are both harmonic oscillator potentials.
Because there is only a translation between the two, one may believe
that both are admissible. This is however not the case using the
general method of Ref.~\cite{Ref5}. Using the different
functions $W_L, L=0,1,2$ which enter \Eqs{Eq26}--(\ref{Eq32}), the case
which would lead to $V(x)$ implies unbounded solutions \Eq{Eq26}, \Eq{Eq28} 
and \Eq{Eq29} at infinity. Only the case which lead to the potential
$\tilde{V}(x)$ has $W_L$-functions which ensure that the eigenfunctions are
physical. Moreover, changing minimally the problem leads often
to the impossibility to solve it. For example, taking  
$U(x)= 3 x^2$, corresponding with non equally spaced
energy levels $E_0=0,E_1=3,E_2=4$, leads to well-defined potentials,
because $W_0(x)$ behaves adequately.
It is however impossible to find physically acceptable wave-functions
due to singularities and bad behavior at infinity of $W_1(x)$ and $W_2(x)$.

Another remark concerns the class of functions given in Ref.~\cite{Ref8},
which is too restrictive. For example, the well-known \cite{Ref3} sextic
oscillator 
\be
V_0(x)=x^6-11x^2+8
\label{Eq46}
\ee
admitting the three solutions
\beqn
\psi_0(x)&=&(1+4x^2+2x^4)\exp\left(-\frac{1}{4}x^4\right), \; E_0=0
\label{Eq47}\\
\psi_1(x)&=&(2x^4-3)\exp\left(-\frac{1}{4}x^4\right), \; E_1=8
\label{Eq48}\\
\psi_2(x)&=&(1-4x^2+2x^4)\exp\left(-\frac{1}{4}x^4\right), \; E_2=16.
\label{Eq49}
\eeqn
corresponds to
\be
U(x)=\frac{(2x^4-3)(2x^4-1)}{x^2}.
\label{Eq50}
\ee
This $U(x)$ does not enter the Kuliy and Tkachuk formalism.

We have also tried different other choices for $U(x)$, which do not
belong to the class of   Kuliy and Tkachuk.
We have investigated the two simplest deviations of the quadratic choice:
$U(x)=c (x^2-a^2)$ and $U(x)=c x^2(x^2-a^2)$, with both
choices $a^2>0$ or $a^2<0$. The result is that it is possible to find 
solutions for the first case. These solutions are however not physical
(they diverge at infinity).
In the second case, it is possible to find physically acceptable
solutions. However, it is not possible to express them
analytically because the functions $W_L, W_{L+1}$ and $W_{L+2}$ which enter 
\Eq{Eq26} and \Eqs{Eq28}--(\ref{Eq29}) are not analytically integrable.

\section{Conclusions}
\label{Section5}

In this paper, we have presented a powerful alternative to the method
of Kuliy and Tkachuk of Ref.~\cite{Ref5}. We have shown that the function
$U(x)$ is, even in a case which differs minimally from the one
treated in that reference, complicated and heavy to handle: simple $U(x)$
do not lead to Q.E.S. potential; other $U(x)$ may give a good potential,
but not physically relevant solutions.
In 
contradistinction, our method seems straightforward to apply and,
by construction, leads to physical solutions.

\end{document}